\documentclass[]{article}
\newcommand{\SU}{\mathrm{SU}}
\newcommand{\Ref}[1]{(\ref{#1})}

\let\eps=\epsilon

\newcommand{\eqa}{\begin{eqnarray}}
\newcommand{\neqa}{\end{eqnarray}}
\newcommand{\equ}{\begin{equation}}
\newcommand{\nequ}{\end{equation}}
\newcommand{\no}{\nonumber\\}
\def\f{\frac}
\def\arr{\rightarrow}
\newcommand{\hh}{{\cal H}}
\newcommand{\I}{{\cal I}}
\newcommand{\te}[1]{\widehat{\theta}_{#1}}
\newcommand{\ket}[1]{|{#1}\ra}
\def\ra{\rangle}
\def\la{\langle}
\newcommand{\6}{$\{6j\}$}
\def\d{\delta}
\newcommand{\mean}[1]{\la{#1}\ra}
\newcommand{\bra}[1]{\la {#1}|}
\newcommand{\p}{\partial}
\newcommand{\lp}{\ell_{\rm P}}
\usepackage{bbm}

\begin{document}
\title{\LARGE \bf A semiclassical tetrahedron} 
\author{Carlo Rovelli 
and Simone Speziale\footnote{rovelli@cpt.univ-mrs.fr, sspeziale@perimeterinstitute.ca}\\[2mm]
\normalsize \em CPT%
\footnote{Unit\'e mixte de recherche (UMR 6207) du CNRS et des Universit\'es
de Provence (Aix-Marseille I), de la Meditarran\'ee (Aix-Marseille II) et du Sud (Toulon-Var); laboratoire affili\'e \`a la FRUMAM (FR 2291).} , CNRS Case 907, Universit\'e de la M\'editerran\'ee, F-13288 Marseille\\
\normalsize \em Perimeter Institute, 31 Caroline St.N, Waterloo, ON-N2L-2Y5,
Canada}
\date{\small\today}
\maketitle\vspace{-7mm}
\begin{abstract}
\noindent We construct a macroscopic semiclassical state state for a quantum tetrahedron. 
The expectation values of the geometrical operators representing the volume, areas and dihedral angles
are peaked around assigned classical values, with vanishing relative uncertainties.
\end{abstract}

\section{Introduction}

In loop quantum gravity (LQG), the geometry of the physical space turns out to be quantised \cite{eigen,eigen2}.  In particular, by studying the spectral problem associated to the operators representing geometrical quantities, one finds two families of quantum numbers, which have a direct geometrical interpretation:  $\SU(2)$ spins, labeling the links of a spin network, and $\SU(2)$ intertwiners, labeling its nodes.  The spins are associated to the area of surfaces intersected by the link, while the intertwiners are associated to the volume of spatial regions that include the node, and to the angles formed by surfaces intersected by the links (see \cite{Major:2001zg}).
A four-valent link, for instance, can be interpreted as a ``quantum tetrahedron": an elementary ``atom of space" whose face areas, volume and dihedral angles are determined by the spin and intertwiner quantum numbers. See for instance \cite{book} for a detailed introduction and full references. 

Remarkably, the very same geometrical interpretation for spins and intertwiners can be obtained from a formal quantisation of the degrees of freedom of the geometry of a tetrahedron \cite{Barbieri, BaezTet}, without any reference to the full quantisation of general relativity which is at the base of LQG.  In this case, one can directly obtain the Hilbert space $\cal H$ describing a single quantum tetrahedron. The states in $\cal H$ can be interpreted as ``quantum states of a tetrahedron", and the resulting quantum geometry is the same as the one defined by LQG. 
 
In this quantum geometry, not all the variables describing the geometry of the tetrahedron turn out to commute.  Consequently, in general there is no state in $\cal H$ that corresponds to a given classical geometry of the tetrahedron.  This fact raises immediately the problem of finding semiclassical quantum states in $\cal H$ that approximate a given classical geometry, in the sense in which wave packets or coherent states approximate classical configurations in ordinary quantum theory.  This is the problem of defining the ``coherent tetrahedron".   
The problem of constructing coherent states in LQG has raised an increasing interest over the last few years \cite{Bombelli, Bombelli2, Bombelli3}, in particular in relation to the possibility of studying the low energy limit of LQG, which is one of the main open issues in this approach to quantum gravity.  For instance, writing semiclassical tetrahedron states is needed in order to develop the program for computing $n$-point functions in LQG initiated in \cite{RovelliProp,Io,Bianchi:2006uf,Livine}. 
  
In this paper we propose an explicit construction of a semiclassical quantum state, corresponding to a given macroscopic geometry of the tetrahedron. 

\section{Quantum geometry of the tetrahedron}\label{quantumtet}
Let us first summarise well known facts about the quantum geometry of an atom of space.  For simplicity, we do not refer to full LQG, but rather to the direct quantisation of the degrees of freedom of a tetrahedron.

Consider four irreducible representations (irreps) $j_i$, with $i=1, ...4$, of $\SU(2)$. Let  ${\cal H}_{j_i}$ be the corresponding representation spaces.  The tensor product of these four spaces carries a reducible representation of $\SU(2)$, that can be decomposed in its irreducible components.  Denote the ensemble of the spin-zero components, namely the $\SU(2)$ invariant component of the tensor product as 
\equ\label{Inv}
{\cal I}_{j_1\ldots j_4} := {\rm Inv} \left[ \bigotimes_{i=1}^4 {\cal H}_{j_i}\right]
\nequ
As we show below following \cite{Barbieri, BaezTet}, this space 
can be interpreted as the space of the quantum states of a quantum tetrahedron whose 
$i$-th triangle has area given by the (square root of the) $\SU(2)$ Casimir operator, $A_i=\lp^2 C(j_i)$. 
In the following we work in units $\lp=1$, and we take $C^2(j)=(j+\f12)^2$.

The Hilbert space ${\cal H}:=\bigoplus_{j_i}{\cal I}_{j_1\ldots j_4}$ describes the degrees of freedom associated to the volume and the dihedral angles of this atom of quantum geometry. 
Let us see how this Hilbert space is related to the classical geometry of a tetrahedron. 
The classical geometry of a tetrahedron, modulo rotations and
translations, is fully determined by six parameters, for instance the lengths of its six sides, or the area of its four triangles and two dihedral angles between these faces.  This latter case is suitable for comparison with the quantum theory.
Let us call $\vec {n}_i$, $i=1,..,4$ the four normals to the triangles pointing outward, with length determined by the triangle area as
$|\vec{n}_i|\equiv 2A_i$.  The
dihedral angles $\theta_{ij}$ are given by the scalar products $\vec
n_i \cdot \vec n_j = |\vec{n}_i| |\vec{n}_j| \cos\theta_{ij}$, $i\neq j$.
There are relation between the variables $\vec{n}_i, A_i$. 
First, we have the closure constraint $\sum_{i=1...4}\vec{n}_i=0$.  Second, for any two opposite angles we have a relation of the form $
\vec n_3 \cdot \vec n_4=\vec n_1 \cdot \vec n_2+\left(A_1+A_2-A_3-A_4\right)$.
In terms of the $\vec n_i$, the volume of the tetrahedron is given by the simple relation:
\equ
\label{volumetet}
V^2 = -\frac{1}{36}\ \eps_{abc}\ n_1^a n_2^b n_3^c = -\frac{1}{36} \ \vec n_1 \cdot \vec n_2 \times \vec n_3.
\nequ
The geometry of the tetrahedron is thus completely determined, for instance by the variables 
\linebreak $A_1,..,A_4, \theta_{12},\theta_{13}$.

The geometric quantisation of these degrees of freedom is based on the identification of  generators of $\SU(2)$ as quantum operators corresponding to the $\vec{n}_i$ \cite{BaezTet}. 
As mentioned, this construction gives directly the same quantum geometry that one finds via a much longer path by quantising the phase space of general relativity. 
The squared lengths $|\vec{n}_i|^2$ are the $\SU(2)$ Casimirs $C^2(j)$, as in LQG. A quantum state of a tetrahedron with fixed values of the area must therefore live in the tensor product  $\bigotimes_{i=1}^4 {\cal H}_{j_i}$of the spin $j_i$ representations spaces. The closure constraint now reads:
\equ
\label{closure}
\sum_{i=1}^4 \vec J_i = 0,
\nequ
and imposes that the state of the quantum tetrahedron is invariant under
global rotations (simultaneous $\SU(2)$ rotations of the four
triangles). Therefore it is a singlet state, namely an {intertwiner} map $\bigotimes_{i=1}^4 {\cal
 H}_{j_i}\arr\hh_{j=0}\equiv{\mathbbm C}$. The state space of the quantum tetrahedron with given areas is thus the Hilbert space of intertwiners $\I_{j_1..j_4}$ given in \Ref{Inv}. The operators $J_i{}^2$, $\vec J_i \cdot \vec J_j$ are well defined on this space, and so is the operator,
\equ\label{defU}
U:= -\epsilon_{abc} J_1^aJ_2^bJ_3^c.
\nequ
$U$ has a symmetric positive/negative
spectrum: if $u$ is an eigenvalue, so is $-u$ (see \cite{Barbieri, DePietriRecoupling}). Its absolute value $|U|$ can immediately be identified with the quantisation of the
classical squared volume $36V^2$, by analogy with \Ref{volumetet}, again in agreement with standard LQG results. 

To find the angle operators, let us introduce the quantities $\vec{J}_{ij}:=\vec J_i +\vec J_j$.
Their geometrical interpretation can be found applying the same arguments as above to $\vec n_i + \vec n_j$. 
It turns out that $\sqrt{{J}_{ij}{}^2}$ is proportional to the area $A_{ij}$ of the internal parallelogram, whose 
vertices are given by the midpoints of the segments belonging to either the triangle $i$ 
or the triangle $j$ but not to both (see \cite{BaezTet}),
$A_{ij}:= \f{1}{4\sqrt{2}}\sqrt{{J}_{ij}{}^2}$.
Given these quantities, the angle operators $\te{ij}$ can be recovered from
\equ\label{angle}
J_i \, J_j \, {\cos\te{ij}} = \vec J_i\cdot \vec J_j = \f12(J_{ij}^2-J_i^2 -J_j^2).
\nequ

We conclude that the quantum geometry of a tetrahedron is encoded in the operators
$J_i^2, J_{ij}^2$, $U$, acting on $\cal H$.
It is a fact that out of the six independent classical variables, only
five commute in the quantum theory. Indeed while we have
$[{J}_k{}^2, J_i \cdot  J_j]=0$, it is easy to see that: 
\equ
\left[  J_1\cdot  J_2\, , J_1\cdot J_3  \right] 
\,=\, \f14 \left[J_{12}^2, J_{13}^2 \right]\,=\,
i\, \epsilon_{abc}
\,J_1^aJ_2^bJ_3^c \equiv -i U \ne 0.
\label{thecommutator}
\nequ
A complete set of commuting operators, in the sense of Dirac, is given by 
the operators $\{J^2_i, J^2_{12}\}$. In other words, 
a basis for $\I_{j_1\ldots j_4}$ is provided by the
eigenvectors of any one of the operators $J^2_{ij}$.  We write the 
corresponding eigenbasis as $\ket{j}_{ij}$.  For instance, 
the basis $\ket{j}_{12}$ diagonalises the four triangle areas and the dihedral angle
$\theta_{12}$ (or, equivalently, the area $A_{12}$ of one internal parallelogram).

The relation between different basis is easily obtained from $\SU(2)$ recoupling theory: 
the matrix describing the change of basis in
the space of intertwiners is given by the usual Wigner $\{6j \}$ symbol,
\equ
\label{6j}
W_{jk}:= {}_{12}\la j | k\ra_{13} = (-1)^{\sum_i j_i} \sqrt{d_j\,d_k}
\left\{ \begin{array}{ccc} j_1 & j_2 & j \\ j_3 & j_4 & k \end{array} \right\},
\nequ
so that
\equ\label{rec}
\ket{k}_{13}=\sum_j W_{jk} \ket{j}_{12}.
\nequ
Here we used the notation $d_j=2j+1$.
Notice that from the orthogonality relation of the \6 symbol,
\equ
\sum_i d_i \left\{ \begin{array}{ccc} j_1 & j_2 & i \\ j_3 & j_4 & j \end{array} \right\}
\left\{ \begin{array}{ccc} j_1 & j_2 & i \\ j_3 & j_4 & k \end{array} \right\} = 
\f{\d_{jk}}{d_j},
\nequ
we have
\equ\label{ortho}
\sum_i W_{ij} W_{ik} = \d_{jk}.
\nequ

The states $\ket{j}_{12}$ are
eigenvectors of the five commuting geometrical operators $\{J^2_i, J^2_{12}\}$, thus the average
value of the operator corresponding to the sixth classical observable, say $J^2_{13}$,
is on these states maximally spread.   This means
that a basis state has undetermined classical geometry or, 
in other words, is not an eigenstate of the geometry. 
We are then led to consider superpositions of states to be able 
to study the semiclassical  limit of the geometry. 
Suitable superpositions could be constructed for instance requiring that they
minimise the uncertainty relations between non--commuting  observables,
such as
\equ\label{unce}
\Delta^2 J_{12}^2\, \Delta^2 J_{13}^2 \geq \f14 | \mean{[J_{12}^2, J_{13}^2]}|^2 \equiv 4 \, |\mean{U}|^2.
\nequ
States minimising the uncertainty above are usually called coherent states.

In principle one has two options, (i) to work
within the space $\I_{j_1\ldots j_4}$, namely at fixed values of the spins, or 
(ii) to work in the whole space ${\cal H}$. In the first case one is interested in
semiclassical states with sharp values of the triangle areas and fuzzy values of the dihedral angles;
in the second case one considers also the possibility of fuzzy values of the external areas.
Here we consider the first option, and we show below how to construct states in $\I_{j_1\ldots j_4}$
such that all relative uncertainties ${\mean{\Delta^2 J_{ij}}}/{\mean{J^2_{ij}}}$, or equivalently
${\mean{\Delta \te{ij}}}/{\mean{\te{ij}}}$, vanish in the large scale
limit. The latter is defined by taking the limit when all spins involved go 
uniformly to infinity, namely $j_i = n k_i$ with $n\mapsto\infty$.

Notice that this is a different requirement than minimising \Ref{unce}, thus we expect the semiclassical states
constructed here not to be coherent states.

\section{Semiclassical states}

To fix ideas, we choose a classical geometry $A_1,\ldots A_4,\theta_{12},\theta_{13}$, and we 
work in the basis $\ket{j}_{12}$. Let us consider a generic state
\equ\label{psi1}
\ket{\psi}  = \sum_j c_j \ket{j}_{12} \in \I_{j_1\ldots j_4}.
\nequ
We want to select the coefficients $c_j$ such that 
\equ\label{pluto}
\mean{\te{ij}}\mapsto\theta_{ij}, \qquad \f{\mean{\Delta\te{ij}}}{\mean{\te{ij}}}\mapsto 0
\nequ
in the large scale limit, for all $ij$.
The large scale limit considered here is taken when all spins are large. Consequently, in the 
following we approximate $j+\f12\sim j$.

\subsection{Gaussians around $\theta_{12}$}
We begin by considering the expectation value of $J_{12}^2$, 
\equ\label{j12}
\f{\bra{\psi} J_{12}^2 \ket{\psi}}{\bra\psi\psi\ra} = \f{\sum_j |c_j|^2 \, C^2(j)}{\sum_j |c_j|^2 },
\nequ
from which we can study the
angle operator $\te{12}$ using \Ref{angle}. We can easily peak the expectation value of $J_{12}^2$ using
a Gaussian distribution in \Ref{psi1}, such as
\equ\label{c1}
c_j(j_0) = \f1{\sqrt[4]{2\pi\sigma_j}}\,\exp\left\{-\f{(j-j_0)^2}{4\sigma_j}\right\}.
\nequ
Here $j_0$ is a given real number to be linked to $\theta_{12}$ below. We allow the variance
$\sigma_j$ to have a dependence on $j_0$, but we restrict this dependence to be of the type $j_0^p$
with $p<2$. More precisely, using the scale parameter
$n$ introduced above, this condition reads $\sigma_j \propto n^p$ with $p<2$.
In the large $j$ regime, we can approximate the sum in \Ref{j12} with an integral, 
$\sum_j \sim \int_{j_{\rm min}}^{j_{\rm max}} dj \sim \int_{-\infty}^\infty d\d j$, 
where $\d j=j-j_0$, and we can compute
\eqa\label{2}
\mean{J_{12}} &\simeq& \f1{\sqrt{2\pi\sigma_j}} \int d \d j \, e^{-\f{(j-j_0)^2}{2\sigma_j}} \, 
(j+\f12) = C(j_0), \\ \label{1}
\mean{J_{12}^2} &\simeq& \f1{\sqrt{2\pi\sigma_j}} \int d \d j \, e^{-\f{(j-j_0)^2}{2\sigma_j}} \, 
(j+\f12)^2 = C^2(j_0)+\sigma_j,
\neqa
so that $\mean{\Delta^2 J_{12}} = \sigma_j$. With the above assumption on the $j_0$ dependence of 
$\sigma_j$, we have 
\equ\label{limit}
\f{\mean{\Delta^2 J_{12}}}{\mean{J_{12}^2}}\simeq \f{\sigma_j}{j_0^2}\mapsto 0
\nequ
in the limit $j_0\mapsto\infty$.
The expectation value of $J_{12}$
is peaked around $j_0$, with vanishing relative uncertainty.
Consequently, also the angle operator will be peaked, 
\equ\label{t12a}
\mean{\cos\te{12}}\simeq \f{j_0^2-j_1^2-j_2^2}{2 j_1 j_2}.
\nequ
Using this expression for the expectation value of the angle operator, it is easy to
express the parameter $j_0$ as a function of the desired classical value $\theta_{12}$,
\equ\label{t12}
j_0^2=2j_1j_2\cos\theta_{12}+j_1^2+j_2^2.
\nequ

\subsection{Phases around $\theta_{13}$: the auxiliary tetrahedron}

The next step is to modify \Ref{c1} such that
also $\widehat\theta_{13}$ is peaked around the classical value $\theta_{13}$, with vanishing relative uncertainty.
To do so, notice that the results obtained above for $\te{12}$ do not change if we add a phase to \Ref{c1}.
To understand what is the right phase to add to peak $\te{13}$,
let us inspect the transformation property \Ref{rec}. This is mainly determined by the \6 symbol.
Now, we know from the Ponzano--Regge model for 3d quantum gravity that the \6 symbol is the quantum amplitude
of a tetrahedron whose edge lengths are given by  the (Casimirs of the) six half--integers entering the symbol.
Then, let us consider an auxiliary tetrahedron, whose six edge lengths are given by 
$j_1, \ldots j_4, j_0, k_0$, where $j_0$ is given by (the biggest half--integer smaller than) \Ref{t12} and
$k_0$ is a (similar) function of $\theta_{13}$ to be computed below.
In constructing the auxiliary tetrahedron, we take $j_0$ and $k_0$ to be opposite edges, and $j_1, j_2$ and $j_0$ to share a vertex. Consequently, $j_1$ $j_2$ and $k_0$ are coplanar.
{} From the edge lengths, we can compute the dihedral angles of this auxiliary tetrahedron. In 
particular, let us consider the dihedral angles to $j_0$ and $k_0$, which
we call $\phi(j_0, k_0)$ and $\chi(j_0, k_0)$ (we omit, for brevity, the dependence on the fixed $j_1\ldots j_4$.). 
They can be computed from the well--known formulae
\equ\label{dihedral}
\sin\phi(j_0, k_0) = \f32 \f{(j_0+\f12) V(j_e)}{A_1 A_2}, \qquad 
\sin\chi(j_0, k_0) = \f32 \f{(k_0+\f12) V(j_e)}{A_3 A_4},
\nequ
where $V(j_e)$ is the volume of the tetrahedron with edge lengths $\ell_e=j_e+\f12$ 
and $A_1$, $A_2$ (respectively $A_3$, $A_4$) are the triangles
sharing the edge $j_0$ ($k_0$). Here we introduced the notation $j_e=\{j_i, j_0, k_0\}$.

We now consider the state \Ref{psi1} with the following coefficients,
\equ\label{c2}
c_{j}(j_0, k_0) = \f1{\sqrt[4]{2\pi\sigma_j}} \, \exp\left\{-\f{(j-j_0)^2}{4\sigma_j} + i\phi(j_0, k_0) j\right\}.
\nequ
For the moment, we still do not fix the value of the variance $\sigma_j$. As we simply added a phase to \Ref{c1},
this new state still guarantees \Ref{1} and \Ref{2}. Let us study the expectation value of $J_{13}^2$.
Using \Ref{rec}, we can write
\equ
\ket{\psi}=\sum_{k} c'_k(j_0, k_0) \ket{k}_{13}, \nequ
with
\equ\label{c'} 
c'_k(j_0, k_0)=\sum_j c_j W_{jk}.
\nequ
We have straightforwardly
\equ
\bra{\psi}J_{13}^2\ket{\psi}=\sum_k |c_k'|^2 C^2(k),
\nequ
thus the expectation value of $J_{13}$ as well as its uncertainty are
determined by the coefficients $c_k'$. Their exact evaluation is rather non trivial.
However, since we are interested only in the large scale limit,
we compute the $c_k'$ only for large spins. The expansion in the spins of course spoils the normalisation,
thus in the following we will consider the normalised expectation value 
$\bra{\psi}J_{13}^2\ket{\psi}/\bra{\psi}{\psi}\ra$.

To study the large $j$ expansion of \Ref{c'}, we can use the well--known formula
for the asymptotics of the \6 symbol \cite{Ponzano, asympt, asympt2, asympt3},
\equ\label{asymp}
\left\{ \begin{array}{ccc} j_1 & j_2 & j \\ j_3 & j_4 & k \end{array} \right\}  
\simeq \f{\cos\left( S_{\rm R}[j_e] +\f\pi 4  \right)}{\sqrt{12\, \pi\, V(j_e)}},
\nequ
where $S_{\rm R}[j_e]$ is the Regge action of the auxiliary tetrahedron, 
\equ\label{regge}
S_{\rm R}[j_e] = \sum_{e}\, (j_e+\f12) \, \phi_e(j_e),
\nequ
and the $\phi_e$ are the dihedral angles, whose expressions in terms of edge lengths 
are as in \Ref{dihedral}.
Using \Ref{asymp}, we can write \Ref{c'} as
\equ\label{cregge}
c_k' \simeq \sum_j \f{\mu(j,k)}{\sqrt[4]{2\pi\sigma_j}}
\cos\left( S_{\rm R}[j_e]+\f\pi4\right)
\exp\left\{ -\f{(j-j_0)^2}{4\sigma_j}+i\phi(j_0, k_0) j\right\},
\nequ
with 
\equ\label{measure}
\mu(j,k)=\sqrt{\f{d_j d_k}{12\, \pi\, V(j_i, j, k)}}.
\nequ
Recall that the Regge action is a discretised version of GR, which captures the non--linearity of the theory.
Because of the Gaussian in \Ref{cregge}, we can expand the Regge action and $\mu(j, k)$
around the values $j=j_0$, $k=k_0$. Denoting $\d j=j-j_0$, $\d k = k-k_0$, we have
\eqa\label{Regge1}
S_{\rm R}[j_e] &=& S_{\rm R}[j_0, k_0] +
\f{\p S_{\rm R}}{\p j}{\Big|_{j_0, k_0}} \d j + \f{\p S_{\rm R}}{\p k}{\Big|_{j_0, k_0}} \d k +\ldots =
\no &=&S_0[j_i] + \phi(j_0, k_0)  j + \chi(j_0, k_0)  k + \f{1}{2} G_{jj} \d j^2
+\f12 G_{kk} \d k^2 + G_{jk} \d j \d k +\ldots,
\neqa
where $S_0[j_i]=\sum_{i=1}^4 (j_i+\f12)\phi_i(j_0, k_0)$, and we have introduced the shorthand notation 
$$G_{jj} = \f{\p^2 S_{\rm R}}{\p j^2}{\Big|_{j_0, k_0}}, \qquad
G_{kk} = \f{\p^2 S_{\rm R}}{\p k^2}{\Big|_{j_0, k_0}}, \qquad
G_{jk} = \f{\p^2 S_{\rm R}}{\p j \p k}{\Big|_{j_0, k_0}}.
$$
These coefficients can be evaluated from elementary geometry, using the formulae \Ref{dihedral}
for the dihedral angles (see for instance the Appendix of \cite{Livine}). By dimensional analysis
it follows that $G\sim 1/j$.

Notice the term $\phi(j_0, k_0) j$ appearing in \Ref{Regge1}: this is the phase of the Gaussian in \Ref{cregge}. 
Therefore, when we use \Ref{Regge1} in \Ref{cregge}, this phase
is cancelled or doubled, depending on the sign
of the two exponentials of the cosine. But because the phase makes the argument of the sum
rapidly oscillating, we expect only the exponential where the phase is cancelled 
to contribute to the sum. This mechanism was first noted in \cite{RovelliProp}, and
numerically confirmed in \cite{Io}.

From the analysis of \cite{Livine}, we know that only the background value $\mu(j_0, k_0)$
enters the leading order of \Ref{cregge}. We can thus write simply
\equ\label{c'1}
c_k' \simeq  \f{\mu(j_0, k_0)}{2\sqrt[4]{2\pi\sigma_j}}
\exp\left\{-i S_0[j_i] -i\chi(j_0, k_0) k  \right\}
\sum_j \exp\left\{-\f12(\f1{2\sigma_j}+iG_{jj}){\d j^2}
-i G_{jk} \d j \d k -\f i2G_{kk} \d k^2 \right\}.
\nequ
The factor $S_0[j_i]$ gives an irrelevant global phase, and we disregard it in the following.
This sum can be computed approximating it with an integral as we did above, and we obtain
\equ
c_k' \simeq \f{\mu(j_0, k_0)}{2\sqrt[4]{2\pi\sigma_j}}
\sqrt{\f{2\pi}{\f1{2\sigma_j} +i G_{jj}}}
\exp\left\{-\f12\left(\f{G_{jk}^2}{(\f1{2\sigma_j} +i G_{jj})} + iG_{kk} \right)\d k^2 -i\chi(j_0, k_0) k   \right\}.
\nequ
We have obtained a Gaussian distribution in $k$, with variance
\equ\label{sk}
\sigma_k := \f12\left(\f{G_{jk}^2}{\f1{2\sigma_j} +i G_{jj}} + iG_{kk} \right)^{-1}.
\nequ

We can now fix the value of the variance $\sigma_j$, by requiring both $\sigma_j$
and $\sigma_k$ to be real quantities. The imaginary part of \Ref{sk} is
(proportional to) $G_{jj}^2 G_{kk}-G_{jj}G_{jk}^2+\f1{4\sigma_j^2}G_{kk}$,
and imposing it to be zero we obtain the only solution
\equ\label{sj1}
\sigma_j^2=\f{G_{kk}}{4G_{jj}}\f1{G_{jk}^2- G_{jj} G_{kk}},
\nequ
and consequently
\equ\label{sk1}
\sigma_k^2=\f{G_{jj}}{4G_{kk}}\f1{G_{jk}^2- G_{jj} G_{kk}}.
\nequ
The reality of the above variances is guaranteed by the following two 
geometric properties: first, $G_{jj}<0$ and $G_{kk}<0$
due to the monotonic dependence of any dihedral angle on all edge lengths; 
second, $G_{jk}^2-G_{jj} G_{kk}>0$ due to the triangle
inequalities satisfied by the edge lengths.
These properties can be easily verified and we do not provide the proof here.

Notice that because $G\sim 1/j$, we have $\sigma^2\sim j^2$ and thus \Ref{limit} is satisfied with $p=1$.

Using the explicit value \Ref{sk1} we can write \Ref{c'1} as
\equ\label{c'2}
c_k' \simeq \f{N^{(1)}}{\sqrt[4]{2\pi\sigma_k}} \exp\left\{-\f{(k-k_0)^2}{4\sigma_k}-i\chi(j_0, k_0) k \right\},
\nequ
where
$$
N^{(1)}:=\f{\mu(j_0,k_0) \sqrt{\f\pi2}}{\sqrt{\sqrt{G_{jk}^2-G_{jj}G_{kk}}+i\sqrt{G_{jj}G_{kk}}}}
$$
is the correction to the normalisation due to the fact that the coefficients are evaluated only at leading order.

Using \Ref{c'2} in \Ref{psi1} and proceeding as above, it is straightforward to show that 
$$
\f{\bra\psi{J_{13}^2}\ket\psi}{\la\psi|\psi\ra}\simeq C^2(k_0), \qquad
\f{\bra\psi{\Delta^2 J_{13}}\ket\psi}{\la\psi|\psi\ra}\simeq \sigma_k,
$$ 
so that $J_{13}$ is peaked around $k_0$ with vanishing relative uncertainty in the large spin limit.
Then, using
\equ\label{t13a}
\mean{\cos\te{13}}\simeq \f{k_0^2-j_1^2-j_3^2}{2 j_1 j_3},
\nequ
we can link $k_0$ to the classical value $\theta_{13}$,
\equ\label{t13}
k_0^2=2j_1j_3\cos\theta_{13}+j_1^2+j_3^2.
\nequ

This shows that the superposition with coefficients \Ref{c2} is a good semiclassical
state, namely it satisfies \Ref{pluto}.

Notice that the sign of the phase in \Ref{c'2} is opposite to the one in \Ref{c2}; this
can be related to the fact that $j_1, j_2, j$ all belong to the same vertex in the
auxiliary tetrahedron, whereas $j_1, j_2, k$ are coplanar.

\subsection{Equilateral case}
To be more concrete, let us consider a simple example: the equilateral case when
$A_i\equiv A=j$ $\forall i$, $j$ large, and $\theta_{ij}\equiv \theta=\arccos(-\f13)$ $\forall ij$.
{} From the value of $\theta$ we can compute $j_0=k_0=\f2{\sqrt3}j$, using \Ref{t12}.
Notice that the auxiliary tetrahedron is isosceles, not equilateral. The relevant dihedral angle
of the auxiliary tetrahedron can be computed from elementary geometry, and is given by
\equ
\cos{\phi(j_0)}=-\f{{4 j^2-3j_0^2}}{4 j^2-j_0^2}\equiv 0,
\nequ
namely $\phi(j_0)=\f\pi2$. On the other hand,
the $G$ coefficients take the following form \cite{Livine},
\equ
G_{jk} = -\f{\sqrt{2}}{\sqrt{2j^2-j_0^2}}\equiv -\f{\sqrt{3}}{j}, \qquad
G_{jj} = G_{kk} = -\f{\sqrt{2}}{\sqrt{2j^2-j_0^2}}\f{j_0^2}{4j^2-j_0^2}\equiv -\f12\f{\sqrt{3}}{j}.
\nequ
Plugging these values in \Ref{sj1} and \Ref{sk1},
we obtain the very simple variances $\sigma_j = \sigma_k = j_0/3$. 

The state describing an equilateral semiclassical tetrahedron is then
\equ\label{sym}
\ket{\psi}  = \f1{\sqrt[4]{\pi j_0}} \  \sum_j  \  e^{-\f{3}{4j_0}(j-j_0)^2 + i\f\pi2\, j}\ 
\ \ket{j}_{12} \ .
\nequ

\subsection{Volume}
We have shown that the semiclassical state \Ref{c2} encodes the quantities 
$A_1 \ldots A_4, \theta_{12}, \theta_{13}$ as expectation values of geometrical operators.
These values form a complete set of classical observables, thus every other geometrical information can be extracted from them, including the volume of the tetrahedron.
However, it is interesting to see explicitly what the action of the volume operator is on \Ref{c2}.
In Section 2 we introduced the operator $U$ corresponding to the square of the volume. 
To study its action, it is convenient to work in a different basis. Let us introduce the
basis $\ket{u}$ of eigenstates of $U$, 
\equ
U\ket{u} = u \ket{u}, \qquad \ket{u} = \sum_{j}a^u_{j} \, \ket{j}_{12},
\nequ
where the (generalised) recoupling coefficients $a^u_{j}$ satisfy the following recursion relation \cite{mauro},
\eqa
u \, a^u_{j} &=& i \, \alpha_{j+1} \, a^u_{j+1}
-i \, \alpha_{j} \, a^u_{j-1}, \no \no
\alpha_l &=& \f{A({l},j_1+\f12,j_2+\f12)A({l},j_3+\f12,j_4+\f12)}{(2l+1)(2l-1)},
\neqa
where $A(a,b,c)=\f14\Big[(a+b+c)(a+b-c)(a-b+c)(b+c-a) \Big]^{\f12}$ is the area of a triangle with edge lengths $a$, $b$ and $c$.

Using this new basis, it is straightforward to compute
\equ\label{u}
\bra\psi U \ket\psi = \sum_{u}\, |b^u|^2 \, u,
\qquad b^u = \sum_j c_j \, (a^{-1})^u_j.
\nequ
In the large $j$ limit, using the explicit Gaussian expression of \Ref{c1}, we can approximate $b^u\sim a^u_{j_0}$,
so that \Ref{u} reads like the expectation value of the $U$ operator 
in a configuration where $J_{12}^2$ is peaked around the real number $j_0$,
\equ\label{meanU}
\mean{U}\sim \sum_u \, |a^u_{j_0}|^2\, u.
\nequ
Recall that $j_0$ is not an eigenvalue of $J_{12}^2$, so in general $\mean{U}\neq 0$.
We can now refer to the literature on the volume operator
for the semiclassical analysis of \Ref{meanU} (see \cite{eigen, eigen2, Major:2001zg, DePietriRecoupling, Brunnemann:2004xi}).
For instance in the equilateral case with $A=j$, we know from numerical simulations
that \Ref{meanU} with $j_0=\f2{\sqrt3}j$ gives the correct semiclassical value,
$\mean{U}= \f{8}{27\sqrt 3}A^3$. 

We conclude that the state \Ref{c2} is a good semiclassical state for the geometry of a quantum tetrahedron:
the expectation values of the operators are peaked around classical values with vanishing relative uncertainties.
Notice however that it is not a coherent state in $\I_{j_1\ldots j_4}$:
considering the equilateral case for simplicity, it is straightforward to check that we have 
$\Delta^2 J_{12}^2\, \Delta^2 J_{13}^2 \simeq (\f83)^2 j_0^6$, 
whereas $4\,|\mean{U}|^2 = (\f29)^2j_0^6$,
thus the uncertainty \Ref{unce} is not minimised.

\newpage

\section{Conclusions}

Let us summarise the procedure to construct the semiclassical state here proposed:
\begin{itemize}
\item{Choose a classical geometry $A_1\ldots A_4, \theta_{12}, \theta_{13}$, and compute the corresponding
spins via $A_i=j_i$ and $j_0(\theta_{12})$ and $k_0(\theta_{13})$ respectively via \Ref{t12} and \Ref{t13}.}
\item{Pick up an auxiliary tetrahedron with ($\f12$ plus) $j_i, j_0, k_0$ as edge lengths, and compute
the two dihedral angles $\phi$ and $\chi$ via \Ref{dihedral}.}
\item{Choose a basis in $\I_{j_1\ldots j_4}$, say $\ket{j}_{ab}$, and take the linear combination with coefficients
given by \Ref{c2}, namely Gaussians with expectation value the required $j_0$ or $k_0$ and phase given by
the corresponding dihedral angle $\phi$ or $\chi$. The sign of the phase should be opposite
whether the interwiner edge shares or not a vertex with both edges $a$ and $b$ in the auxiliary tetrahedron.}
\end{itemize}
{In the large spin limit, this state satisfies \Ref{pluto}.} Therefore, it encodes the classical values
of the chosen geometry.

\subsection*{Acknowledgments}
We would like to thank Etera Livine for useful comments and suggestions.
We are grateful to our referee for pointing out a number of imprecisions.

\end{document}